\documentstyle[epsfig, 12pt]{article}
\begin{document}
\baselineskip16pt
\def\pslash{\rlap{\hspace{0.02cm}/}{p}}
\def\beq{\begin{eqnarray}}
\def\eeq{\end  {eqnarray}}
\def\P{{\cal P}}
\def\pr{^{\prime}}
\def\GeV{{\rm GeV} }
\def\ln{{\rm ln}}
\def\non{\nonumber}
\def\dirac#1{#1\llap{/}}
\def\as{\alpha_s}
\def\pv#1{\vec{#1}_\perp}
\def\lqcd{\Lambda_{\rm QCD}}

\newcommand\epjc[3]{Eur.\ Phys.\ J.\ C {\bf #1} (#2) #3}
\newcommand\ijmpa[3]{Int.\ J.\ Mod.\ Phys.\ A {\bf #1} (#2) #3}
\newcommand\jhep[3]{J.\ High Ener.\ Phys.\ {\bf #1} (#2) #3}
\newcommand\npb[3]{Nucl.\ Phys.\ B {\bf #1} (#2) #3}
\newcommand\npps[3]{Nucl.\ Phys.\ B (Proc.\ Suppl.) {\bf #1} (#2) #3}
\newcommand\plb[3]{Phys.\ Lett.\ B {\bf #1} (#2) #3}
\newcommand\prd[3]{Phys.\ Rev.\ D {\bf #1} (#2) #3}
\newcommand\prep[3]{Phys.\ Rep.\ {\bf #1} (#2) #3}
\newcommand\prl[3]{Phys.\ Rev.\ Lett.\ {\bf #1} (#2) #3}
\newcommand\rmp[3]{Rev.\ Mod.\ Phys.\ {\bf #1} (#2) #3}
\newcommand\sjnp[3]{{Sov.\ J.\ Nucl.\ Phys.\ }{\bf#1} (#2) #3}
\newcommand\yf[3]{{Yad.\ Fiz.\ }{\bf#1} (#2) #3}
\newcommand\zpc[3]{Z.\ Phys.\ C {\bf #1} (#2) #3}
\newcommand{\hepph}[1]{{\tt hep-ph/#1}}
\newcommand{\heplat}[1]{{\tt hep-lat/#1}}
\newcommand{\hepex}[1]{{\tt hep-ex/#1}}

\begin{flushright} BIHEP-TH-2002-24 \end{flushright}

\begin{center}
{ \Large\bf ~~\\  ~~\\
The reliability of pQCD approach in pion electromagnetic form
factor}
\end{center}

\vspace{0.5cm} \centerline{ Zheng-Tao Wei $\rm ^{a}$~ and ~
             Mao-Zhi Yang  $\rm ^{b,c}$
 \footnote{e-mail address: weizt@phys.sinica.edu.tw;~~
  yangmz@mail.ihep.ac.cn}}

\vspace{0.5cm}
\begin{center}

$\rm ^{a}$ Institute of Physics, Academia Sinica, Taipei, 115
Taiwan, ROC

$\rm ^{b}$ CCAST(World Laboratory), P.O.Box $8730$, Beijing
$100080$, China

$\rm ^{c}$ Institute of High Energy Physics, P.O.Box $918(4)$,
Beijing $100039$, China
\end{center}

\vspace{1cm}
\begin{abstract}
\vspace{0.2cm} %
\noindent The pion electromagnetic form factor with leading and
next-to-leading twist corrections are studied in the framework of
pQCD approach. We find that, at small momentum transfer regions,
Sudakov effects cannot provide a strong enough suppression of the
non-perturbative contributions coming from large transverse
separations, but at large momentum transfer region,
non-perturbative contributions can be effectively suppressed.  So
pQCD approach can be applied at large momentum transfer region. At
the energy region where experiment can access so far, i.e.
$Q<4\mathrm{GeV}$, pQCD prediction can not be precise because
there may be a quite large part of contributions coming from
non-perturbative region. The calculation of power corrections may
not be performed systematically in pQCD framework.

\end{abstract}

\newpage
\section{Introduction }
There is a general agreement that the standard approach of
perturbative QCD (pQCD) is the correct theory in exclusive
processes in the asymptotic limit $Q\to \infty$ \cite{BLER}.
Although reasonable this theory seems in the ideal world, the test
of this theory is performed only in the pre-asymptotic energy
region. Up to now, the success of pQCD framework in exclusive
process is very limited, such as in pion transition form factor
where there is one hadron involved \cite{Kroll}. The more hadrons
involved in exclusive process, the less prediction power of the
pQCD framework. It is found that the pion electromagnetic (EM)
form factor contains substantial soft endpoint contributions which
destroys the consistence of perturbative method \cite{Isgur}. The
mechanism of Sudakov suppression is introduced to suppress the
soft endpoint contribution  and a modified pQCD formula which
includes Sudakov suppression is given in pion EM form factor
\cite{LiSterman}. It is claimed that pQCD calculation can still be
self-consistent  at about $Q\sim 20\lqcd$ (2\GeV for
$\lqcd=0.1\GeV$).

During the past decade, there is no conceptual development of this
modified pQCD approach. However the applicability of this method
have met many theoretical problems. Recently, this modified pQCD
approach (or say pQCD approach for simplicity) was largely applied
in exclusive B decays to treat the endpoint singularity
\cite{pQCD}. There has been much debate concerning the
applicability of pQCD approach in B system \cite{Sachrajda}. In
\cite{WeiYang}, we investigated the reliability of pQCD approach
in $B\to \pi$ form factors and find that the soft contribution
coming from large transverse separations can be comparable with
the hard contribution. This conclusion should be general for many
exclusive processes, such as pion EM form factor, etc. . From the
analysis in the light-cone sum rule approach, the soft endpoint
contribution is found to be large at the experimentally accessible
energy region \cite{Braun}. So, the study of the reliability of
pQCD approach in EM pion form factor is necessary and important.

It is well-known that the leading order (LO) result for pion form
factor with asymptotic leading-twist distribution amplitude (DA)
is small, about $\frac{1}{4}$ to $\frac{1}{3}$ of the experimental
data \cite{LiSterman, Jakob}. The recent next-to-leading order
(NLO) calculation is given in \cite{Melic}. Their conclusion is
that NLO calculation is reliable only when $Q>5\GeV$. Other
important corrections to the leading-twist LO result are higher
twist corrections. The next-to-leading twist (twist-3 in our case)
contribution is ``chirally enhanced" power correction  and it is
most probably important. The study of twist-3 contribution to
pion EM from factor was performed long ago \cite{twist-3}. The
recent research using pQCD approach is given in \cite{CYH}. Both
of their results found a large twist-3 contribution even larger
than leading twist result at intermediate energy region of $Q\leq
5\GeV$. The large twist-3 contribution seems to destroy twist
expansion. Since there is endpoint singularity in twist-3
contribution, one may also doubt the effectiveness of Sudakov
suppression.

In this paper, we will provide a systematic study of EM pion form
factor in pQCD approach. Our main concerns are the reliability of
applying pQCD method. Some new theoretical ingredients which are
not considered in the previous literatures will be also included
in our analysis: intrinsic transverse momentum effects and
threshold resummation. We find that intrinsic transverse momentum
in pion wave function are important. However, the large twist-3
correction shows that the power corrections may not be
systematically calculated in pQCD framework.

\section{Pion form factor in pQCD approach }

The pion EM  form factor is defined by the following Lorentz
decomposition of biquark current matrix element
\beq
\langle \pi(P^{\prime})|J_{\mu}(0)|\pi(P)\rangle=
 (P+P^{\prime})_{\mu}F_{\pi}(Q^2)
\eeq
where $J_{\mu}=\sum\limits_i e_i \bar q_i \gamma_{\mu}q_i$ is the
electromagnetic current with quark flavor $i$ and relevant
electronic charge $e_i$. The momentum transfer is
$q^2=(P^{\prime}-P)^2=-Q^2$. We have restricted our discussion in
space-like region. It is convenient to use light-cone variables in
which $P=(\frac{Q}{\sqrt 2}, 0, \vec 0_\bot)$ and $P^{\prime}=(0,
\frac{Q}{\sqrt 2}, \vec 0_\bot)$. $F_{\pi}(Q^2)$ is the pion EM
form factor which depends only on momentum transfer $Q$. The pion
form factor at large momentum transfer $Q$ provides information
about the internal structure of pion.

The basic idea of pQCD approach is that it takes into account
transverse momentum and Sudakov suppression. The pion EM form
factor is expressed as the convolution of wave functions $\P$ and
hard scattering kernel $T_H$ by both the longitudinal momentum
faction and the transverse impact parameter $b$: %
\beq \label{eq:pQCD1} %
F_{\pi}(Q^2)=\int dx~ dy~ d^2 \vec b ~d^2 \vec b\pr ~
  \P(x, b, \mu)~ \P(y, b\pr,  \mu) ~
  T_H(x, y, b, b\pr, Q, \mu) . %
\eeq %
The wave function $\P(x, b, Q, \mu)$ is given by: %
\beq \label{eq:P} %
\P(x, b, Q, \mu)={\rm exp}[-S(x, b, Q, \mu)]
 \tilde{\Psi}^0(x,b) %
\eeq %
where $\tilde{\Psi}^0(x, b)$ is the soft part of pion wave
function with $|k_\bot|<1/b$ %
\beq \label{eq:wave} %
\tilde{\Psi}^0(x, b)=\phi(x, 1/b)+O(\as(1/b)) %
\eeq %
The above equation is valid for small $b$. When Sudakov
suppression is strong, there is only small $b$ contribution and
the approximation of $\tilde{\Psi}^0(x, b)$ by distribution
amplitude $\phi(x, 1/b)$ is valid. But at a few $\GeV$ region,
this approximation is questionable.

The factor {\rm exp}(-S) in Eq.(\ref{eq:P}) includes the Sudakov
logarithmic corrections and renormalization group evolution
effects of both wave function and hard kernel, %
\beq %
S(x, b, Q, \mu)=s(x, b, Q)+s(1-x, b, Q)-\frac{1}{\beta_1}
 \ln\frac{\ln(\mu/\lqcd)}{\ln(1/b\lqcd)} %
\eeq %
where $\beta_1=\frac{33-2n_{f}}{12}$ with $n_f=3$. The Sudakov
exponent $s(x, b, Q)$ is calculated up to next-to-leading-log
(NLL) accuracy. Its explicit formula can be found in
\cite{BottsSterman}. The exponent $s(x, b, Q)$ is obtained under
the condition that $xQ/\sqrt 2>1/b$. For small $b$, there is no
suppression, so $s(x, b, Q)$ is set to zero for ${xQ/\sqrt
2}<{1/b}$.


%
%

The study of distribution amplitudes beyond leading twist is
expanded in the conformal spin \cite{Brauntwist}. The light-cone
distribution amplitudes  of pion are defined in terms of bilocal
operator matrix element %
\beq \label{eq:pionwavefunctions} %
\langle \pi(p)|\bar q_\beta(z) q_\alpha|0\rangle
  = \frac{i f_{\pi}}{4} \int_0^1 dx
    e^{i x p\cdot z}
  \times \left\{ \pslash \gamma_5 \phi_{\pi}(x)
    - \mu_{\pi}\gamma_5 \left( \phi_p(x)
    - \sigma_{\mu\nu} p^\mu z^\nu \frac{\phi_\sigma(x)}{6}
    \right) \right\}_{\alpha\beta},  %
\eeq %
where $f_{\pi}$ is the decay constant of pion and $x$ is the
longitudinal momentum fraction of quark in pion. The
parameter $\mu_{\pi}=m_{\pi}^2/(m_u+m_d)$ for charged pion.
$\phi_{\pi}$, $\phi_p$ and $\phi_{\sigma}$ are the twist-2 and
twist-3 distribution amplitudes, respectively. The twist-3 terms
contribute power corrections. At the experimental accessible
energy region, the chirally enhanced parameter
$r_{\pi}=\mu_{\pi}/Q\sim {\cal O}(1)$ is not small.

The momentum projection for pion is \cite{WeiYang}:
\beq \label{eq:pionprojector}
   M_{\alpha\beta}^{\pi} = \frac{i f_{\pi}}{4} \Bigg\{
   \pslash\,\gamma_5\,\phi_{\pi} - \mu_P\gamma_5 \left(
   \phi_p - i\sigma_{\mu\nu}\,\frac{p^\mu\bar p^\nu}{p\cdot\bar p}\,
   \frac{\phi_{\sigma}'}{6}
   + i\sigma_{\mu\nu}\,p^\mu\,\frac{\phi_\sigma}{6}\,
   \frac{\partial}{\partial k_{\perp\nu}} \right)
   \Bigg\}_{\alpha\beta},
\eeq %
where $\phi_{\sigma}'=\frac{\partial \phi_{\sigma} (x)}{\partial
x}$.

The final formula for pion EM form factor in pQCD approach is:
\beq \label{eq:formulae}
F_{\pi}(Q^2)&=& \frac{16}{9}\pi f_{\pi}^2 Q^2
  \int_0^1 dxdy \int_0^{\infty}b db~ b\pr d\pr~ \as(\mu)\times
  \Bigg \{ \frac{\bar y}{2}\P_{\pi}[x]\P_{\pi}[y]
  +\frac{\mu_{\pi}^2}{Q^2} \Bigg [ y\P_p[x]\P_p[y]
  \non \\
 &&  +(1+\bar y)\P_p[x]\frac{\P_{\sigma}\pr[y]}{6}
   +2\P_p[x]\frac{\P_{\sigma}[y]}{6} \Bigg ]
  \Bigg \} \times H(x,y, b, b\pr, Q, \mu)~~~~~~~ \non \\
  &=& F_\pi^{(2)}+F_\pi^{(3)}.
\eeq
where $F_\pi^{(2)}, F_\pi^{(3)}$ represent twist-2 and twist-3
contributions in pQCD approach respectively, and $H$ is given by %
\beq %
H(x,y, b, b\pr, Q, \mu)=
 K_0(\sqrt{\bar x \bar y}~Q b) \Bigg [
   \theta(b-b\pr)K_0(\sqrt{\bar y}Q b)I_0(\sqrt{\bar y}Q b\pr)
   \non \\
  +\theta(b\pr-b)K_0(\sqrt{\bar y}Q b\pr)I_0(\sqrt{\bar y}Q b)
  \Bigg ]~~~~~~ \non
\eeq %
where $[x]=(x, b, Q, \mu)$, $[y]=(y, b\pr, Q, \mu)$. The wave
functions of $\P_{\pi}, \P_p$ and $\P_{\sigma}$ can be obtained
from the relevant wave functions $\Psi_{\pi}, \Psi_p$ and
$\Psi_{\sigma}$ through Eq.(\ref{eq:P}) and Eq.(\ref{eq:wave}).
$K_0$ and $I_0$ are the modified Bessel functions. The choice of
renormalization scale parameter $\mu$ is taken as the largest
momentum scale associated with the exchanged virtual gluon in the
longitudinal and transverse degrees,%
\beq %
\mu={\rm max}(\sqrt{\bar x\bar y}Q, 1/b, 1/b\pr)
\eeq %
The above choice avoids the Landau pole in coupling constant
$\as(\mu)$ at $\mu=\lqcd$ if $\bar x$ and $\bar y$ are small.

In \cite{LiSterman}, the transverse momentum associated with
virtual fermion lines is neglected and the hard kernel involves
only a single impact parameter $b$. In our formula of
Eq.(\ref{eq:formulae}), there are two parameters $b$ and $b\pr$.
The errors caused by the neglect of  transverse momentum  in the
propagator of fermion lines is small if $Q>4\GeV$, but it can be
be about 20\% when $Q=2\GeV$. So, we retain  the transverse
momentum in the fermion lines without approximation. The
transverse momentum $k_\bot^2$ in the numerator are neglected
because it is power suppressed compared to $Q^2$. We checked this
assumption and found its effects are really small in our case.

In the asymptotic limit, the distribution amplitudes
$\phi_{\pi}(x)=6x\bar x$, $\phi_p(x)=1$ and
$\phi_{\sigma}(x)=6x\bar x$. Neglecting the transverse momentum in
both hard kernel and wave functions, Eq.(\ref{eq:formulae})
becomes %
\beq \label{eq:fs}%
F_\pi(Q^2)=\frac{8\pi\as(\mu_{eff}) f_\pi^2}{9Q^2} \left \{
 \left | \int_0^1 \frac{dx}{\bar x}\phi_\pi(x) \right | ^2+
 4\frac{\mu_\pi^2}{Q^2} \left | \int_0^1 \frac{dx}{\bar x}
 \phi_p(x) \right | ^2 \right\}=F_s^{(2)}+F_s^{(3)} %
\eeq %
where $F_s^{(2)}, F_s^{(3)}$ represent twist-2 and twist-3
contributions to $F_\pi(Q^2)$ in the standard approach
respectively. The twist-3 contributes to $1/Q^4$ correction and it
is power suppressed by a factor $\frac{\mu_\pi^2}{Q^2}$ compared
with the leading twist contribution. Because $\phi_p(x)$ is a
constant at endpoint, twist-3 contribution is logarithmically
divergent at $\bar x=0$. We use the effective scale $\mu_{eff}$ in
Eq.(\ref{eq:fs}) because the NLO calculation depends crucially on
the choice of $\mu$.

\section { Numerical results and discussions}

There are only two parameters $\lqcd$ and $\mu_\pi$ in the hard
kernel. They are chosen as $\lqcd=0.2\GeV$ and
$\mu_{\pi}=2.0\GeV$. The distribution amplitudes are taken as
their asymptotic limit form. We do not use C-Z distribution
amplitude for discussion since this model of distribution
amplitude are concentrated at the endpoint where perturbative
analysis is not reliable.

At first, we discuss a problem which is not investigated after the
paper of \cite{LiSterman}: the consistence between the resummed
formula and the standard formula. The modified pQCD approach
uses a b-space resummation formalism in which the Sudakov double
logarithms are resumed to all orders. There is a difficulty of
matching the resumed formalism and the fixed order predictions in
the standard approach. From the theoretical point of view, the
modified pQCD approach should be consistent with standard approach
when $Q$ is large enough. Because of the existence of endpoint
singularity at twist-3 level, twist-3 contribution cannot be
calculated in the standard approach. We can compare the
predictions of pQCD approach and standard approach in leading
twist, i.e., $F_\pi^{(2)}$ and $F_s^{(2)}$ to check the
consistence.

The physical quantity $F_{\pi}$ does not depend on the choice of
the scale parameter $\mu$ if the calculation can be performed up
to infinite orders. However, in practice the calculation can only
be made perturbatively at finite orders. To make the perturbative
expansion reliable, the scale parameter should be chosen in such a
way that make the higher order corrections as small as possible.
In pQCD approach, $\mu={\rm max}(\sqrt{\bar x\bar y}Q, 1/b,
1/b\pr)$. While in the standard approach, it is a free parameter.
We use the different choices proposed in literatures:
$\mu_{eff}=Q, XQ, X^{3/2}Q, e^{-5/6}XQ$ where $X=<\bar x>=1/2$.
Fig.1 shows the $F_s^{(2)}$ in the standard approach with
different choice of $\mu$ and $F_\pi^{(2)}$. One can see that  the
best choice of $\mu$ in the standard approach is $\mu=Q/2$ or
$X^{3/2}Q$. By using this choice, pQCD approach is consistent with
the standard approach when $Q\ge 10\GeV$. For the momentum
transfer $Q<10\GeV$, the prediction of pQCD approach is generally
smaller than that in the standard approach. This difference is
caused by the transverse momentum effects and Sudakov suppression.
In \cite{Melic}, it is found that the choices of $\mu=XQ,
X^{3/2}Q$ reduce the NLO corrections significantly and reliable
predictions can be obtained at lower values of $Q$. Both these
results show the importance of choosing the scale by the interior
dynamics of the process.

\begin{figure}[h]
 \begin{center}
 \epsfig{file=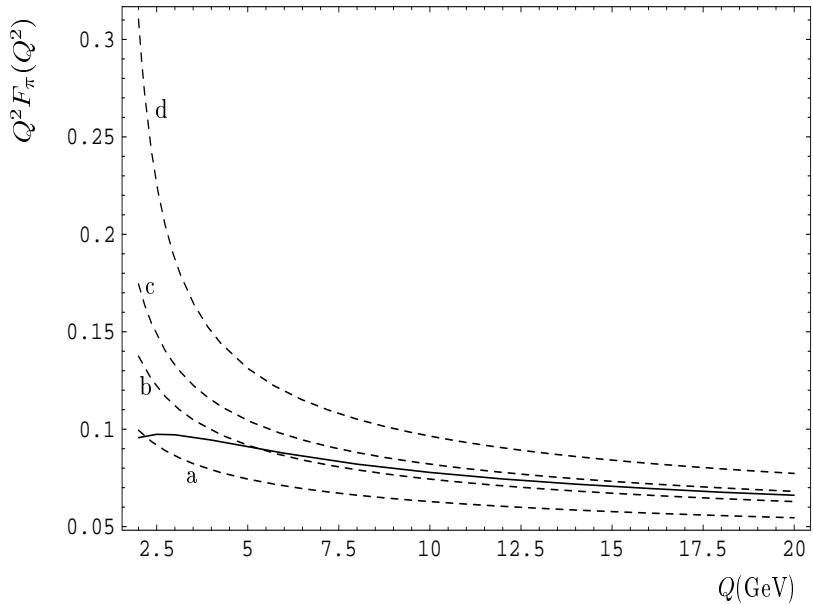, width=12cm,height=10cm}
 \vspace{-1.5cm}
 \end{center}
\caption{ The comparison of twist-2 pion form factor in standard
approach and pQCD approach. The solid line is for pQCD approach,
while the dashed lines for standard approach: (a), (b), (c) and
(d) for $\mu_{eff}=Q$, $XQ$, $X^{3/2}Q$ and $e^{-5/6}XQ$, respectively,
where $X=<\bar x>=1/2$.}
\label{figbpi}
\end{figure}

Now we discuss the reliability of pQCD approach. The basic idea of
pQCD approach is to use Sudakov form factor to suppress the
long-distance contributions coming from large transverse
separations. A reliable perturbative computation should satisfy
that most of the results are coming from small impact parameter $b$. In
order to study the impact parameter $b$ dependence of form factor
$F_{\pi}$, we introduce a cut-off $b^c$ in the impact parameter
space in the integrals of Eq. (\ref{eq:formulae}) by $\int^{b^c}_0 db$.
Similarly, the impact parameter $b\pr$ dependence can also be
performed. It is convenient to use $Q^2F$ as the physical quantity
for discussion because $Q^2F$ is nearly a constant for large Q.
The results are plotted in Fig.2. We can see that
the results are saturated before $b^c$ approaches to
$1/\Lambda_{QCD}$, which means large separation is suppressed by
Sudakov effect. The saturation point is closer to $1/\Lambda_{QCD}$
for smaller momentum transfer. For $Q=10\GeV$ the saturation point
is at about $1.5\GeV^{-1}$, which is far from the end point
$1/\Lambda_{QCD}$. This means that almost all the contributions
come from the short-distance region. For $Q=6\GeV$, the saturation
point is at about $2.5\GeV^{-1}$, which shows that some
non-perturbative contribution emerges, but it is still not large.
But for $Q=4\GeV$ and $2\GeV$, the saturation points are at
$3.5\GeV^{-1}$ and $4.0\GeV^{-1}$, which are quite close to the
end point $1/\Lambda_{QCD}$. There are
substantial contributions coming from large transverse separations
$b >0.5/\lqcd$ for $Q<4\GeV$. Sudakov suppression becomes
weak for small $Q$, and non-perturbative contribution becomes
large. To show directly how the non-perturbative contribution becomes
large at small momentum transfer, we show the $Q^2$ dependence
of $Q^2 F(Q^2)$ in Fig.3. We see that there are quite large
contributions coming from the region $\alpha_s(\mu)>0.5$ for
smaller values of $Q^2$: 34\% at $Q^2=4\GeV ^2$, 22\% at
$Q^2=10\GeV ^2$, 18\% at $Q^2=16\GeV ^2$. For $Q^2>25\GeV ^2$
the contribution for $\alpha_s(\mu)>0.5$ becomes smaller than
10\%.


\begin{figure}[h]
 \begin{center}
 \epsfig{file=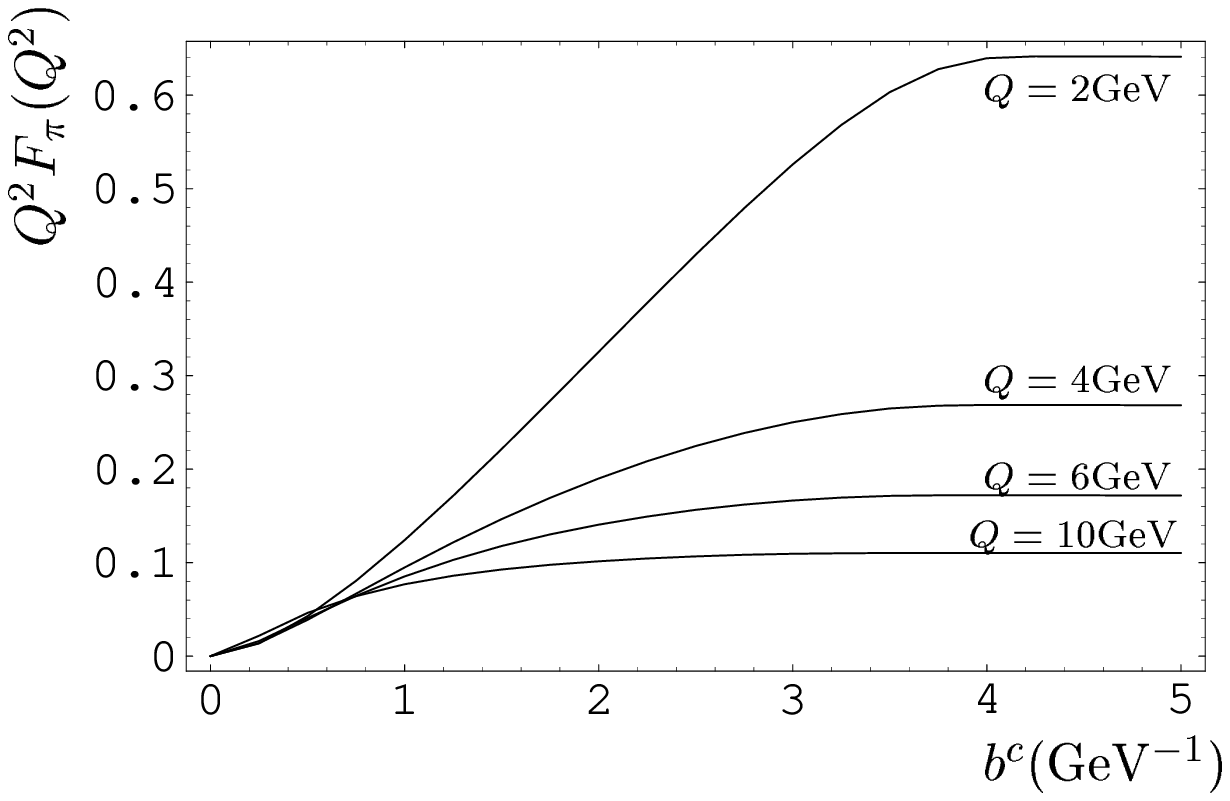, width=10cm,height=8cm}
 \vspace{-1.5cm}
 \end{center}
\caption{ Dependence of $Q^2 F$ on the cut-off $b^c$  at
$Q=2GeV$, $4GeV$, $6GeV$ and $Q=10GeV$. }
\end{figure}

\begin{figure}[h]
 \begin{center}
 \epsfig{file=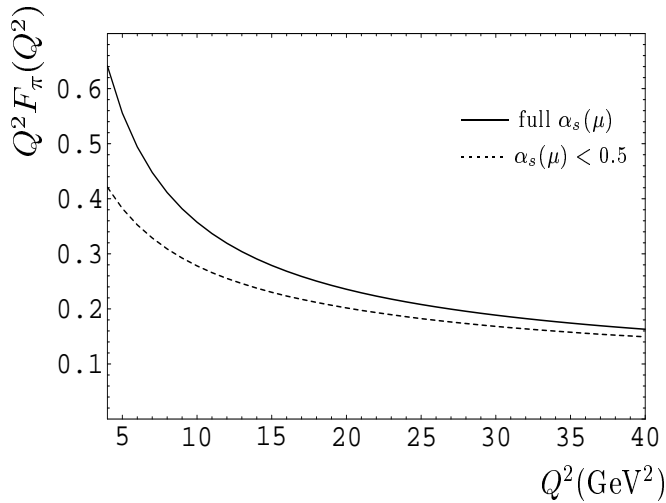, width=10cm,height=8cm}
 \vspace{-1.5cm}
 \end{center}
\caption{ Dependence of $Q^2 F$ on  $Q^2$. The solid
line is the result with the full $\alpha_s(\mu)$, the dashed
line the result with the constraint $\alpha_s(\mu)<0.5$  }
\end{figure}

Is Sudakov suppression effective at experimental accessible energy
region ($Q=1-4\GeV$)? From the above results, it seems that
Sudakov suppression is not strong enough for $Q<4\GeV$. Sudakov
form factor $e^{-s(x,b)-s(\bar x,b)}$ only suppress the region of
$b\sim 1/\lqcd$ at a few $\GeV$ region. In the derivation of
b-space resumed Sudakov form factor, it requires the condition
$Q\gg 1/b \gg \lqcd$. The condition $1/b \gg \lqcd$ is required
because a reliable $\as(1/b)$ expansion needs small $b$. This is
possible only if $Q$ is large enough and Sudakov form factor
suppress all the large $b$ contributions. The derivation of
Sudakov form factor is a difficult problem in QCD because
perturbative expansion is not meaningful at long-distance. This is
unlike the case in QED. Thus, Sudakov form factor depends on the
infrared cut-off. Thus, b-space resummed Sudakov form factor is
not the unique choice. Moreover, the extrapolation of asymptotic
form into the small momentum region is not under well control in
QCD. The predictive power of perturbation theory will be decreased
largely by this extension.
The model-dependent nonperturbative effects have to be included in
order to extrapolate the perturbative analysis to the large $b$
region with the cost that the predictive power of perturbative
method is decreased. In the pion EM form factor, the
non-perurbative contribution at small $Q$ can be at the order of
30\%. This means that Sudakov suppression is not strong enough for
small momentum transfer.

Other mechanisms, such as intrinsic transverse momentum effects
\cite{Jakob} and threshold resummation \cite{Lithreshold} can
suppress large $b$ contribution and endpoint contribution. We
investigate whether these effects can provide a reliable
perturbative analysis. About the intrinsic transverse momentum
effects, it is nonperturbative. The estimate of this effects must
be model-dependent at present. The suppression of the intrinsic
transverse momentum effects is larger than Sudakov effects at the
order of a few $\GeV$. Only for very large $Q$, Sudakov
suppression becomes strong and the intrinsic transverse momentum
effects can be neglected. We incorporate these effects to suppress
the nonperturbative contributions. The formula of including these
effects can be found in \cite{WeiYang}. Here, we do not present
them for simplicity. Fig.4 shows the numerical result after
incorporating these effects. The result is seriously suppressed by
intrinsic transverse momentum effect in pions' wave functions, and
slightly suppressed by threshold resummation, which shows that
intrinsic transverse momentum effect is important in pion EM form
factor. We should include them in our analysis. Fig.5 shows that
the perturbative behavior is improved largely. The saturation
points in $b$-space moves to the $b\to 0$ direction apparently for
both small and large value of $Q$. The contribution of
$\alpha_s(\mu)>0.5$ is largely reduced. For $Q^2>20\GeV ^2$, the
contribution of $\alpha_s(\mu)>0.5$ is suppressed below a few
percent. However, for $Q^2\sim 4\;-\; 10\GeV ^2$ there are 10\% to
20\% of the contributions coming from the region
$\alpha_s(\mu)>0.5$. Non-perturbative contribution for small $Q$
region is still non-negligible. So pQCD can not predict pion EM
form factor precisely in small momentum transfer region. For large
momentum transfer region, the non-perturbative contribution is
small and pQCD can be applied.

\begin{figure}[h]
 \begin{center}
 \epsfig{file=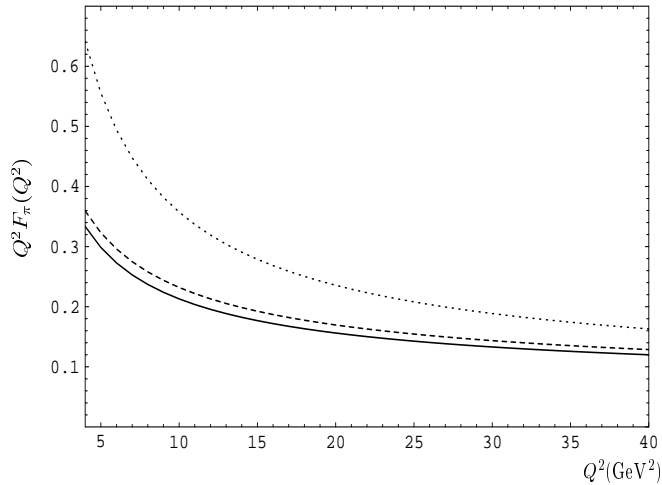, width=10cm,height=8cm}
 \vspace{-1.5cm}
 \end{center}
\caption{ Dependence of $Q^2 F$ on  $Q^2$. The solid
line is the result including intrinsic $k_\perp$ in pion and threshold
resummation effect; the dotted line is without intrinsic $k_\perp$ and
threshold resummation; the dashed line is with intrinsic $k_\perp$ effect
but without threshold resummation.}
\end{figure}

\begin{figure}[h]
 \begin{center}
 \epsfig{file=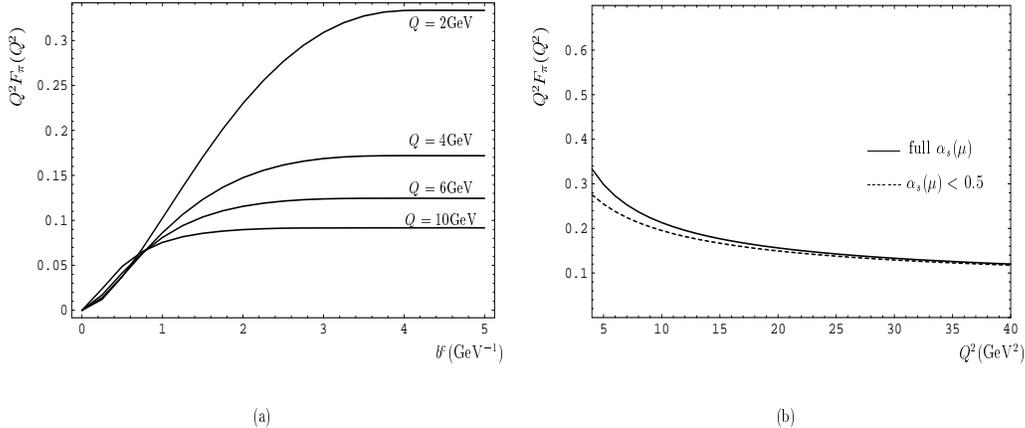, width=15cm,height=8cm}
 \vspace{-1.5cm}
 \end{center}
\caption{ Dependence of $Q^2 F$ on: (a) the cut-off
$b^c$ and (b) $Q^2$, with intrinsic transverse
momentum effects and threshold resummation effects
included.}
\end{figure}

>From the phenomenological point of view, we can use pQCD approach
to estimate the hard contribution. The pQCD approach is valid in
perturbative regions where the strong coupling constant
$\alpha_s(\mu)$ be small. In order to  estimate the hard
contribution, one should make a criterion that a perturbative
contribution should satisfy although it is very hard to define
such a criterion quantitatively. In general perturbative
contribution should come from the region with small strong
coupling constant $\alpha_s$. To make our analysis proceed
numerically, we set a criterion for perturbative contribution:
 $\as<0.5$. This criterion can not be understood as an absolute
 one, it is only indicative. A more stronger criteria may be more
 reliable but may lose some hard contributions.

\begin{table}[hbt]
\begin{center}
\parbox{14cm}{\caption{ The dependence of the hard contribution of pion
 EM form factor $Q^2F$ on $Q$. The rows ``twist-2" and ``twist-3" represent
 twist-2 and twist-3 contributions. The ``total" represents the sum of
 twist-2 and twist-3 contributions}}
\begin{tabular}{|c|c|c|c|c|c|c|c|c|}  \hline \hline
 $Q^2(\GeV^2)$ & 1 & 2 & 4 & 6 & 8 & 10 & 16  &  25   \\\hline
 twist-2  & $0.024$ & 0.042
          & 0.059 & 0.066 & 0.071 & 0.073 & 0.077 & 0.078 \\
 twist-3  & $0.36$ & 0.30 & 0.22 & 0.17 & 0.14 & 0.12
          &  0.086  & 0.060 \\
 total    & $0.38$ & 0.34 & 0.28 & 0.24 & 0.21 & 0.19
          & 0.16  & 0.14 \\ \hline\hline
\end{tabular}
\end{center}
\end{table}

Table 1 gives the numerical results which satisfy $\as<0.5$. From
it, it is seen that the perturbative contribution is about $0.16$
for $Q>4\GeV$ with LO twist-2 and twist-3 corrections. Compared
with the experimental data for $1\GeV ^2<Q^2<6\GeV ^2$, the
results in Table 1 are not too bad.  Considering the large error
bars in the present experimental data, we cannot conclude that
soft contribution in pion EM form factor at small $Q^2$ region can
be neglected. About 20\% percent soft contribution at small
momentum transfer region is still allowed. For $Q^2=1\GeV ^2$, the
soft contributions is larger than 30\%, thus pQCD can not be
applied in this energy region. On the other hand, the results in
Table 1 can not be viewed as the self-consistent prediction of
pQCD because we set $\as<0.5$ in getting them. Only for large
$Q^2$, the results with the constraint $\as<0.5$ can be very near
to the full pQCD prediction. As $Q^2$ becomes large soft
contribution decreases very fast, hard contribution seriously
dominates. Table 1 also shows that contribution of twist-3 is
large for small $Q^2$, it becomes to be smaller as $Q^2$ being
larger.

About power correction, it may contain both hard and soft
contributions from a general point of view. In pion form factor,
we find a large twist-3 contribution which is enhanced by the
endpoint region in pQCD approach. The large twist-3 contribution
means that twist expansion may be questionable. Because Sudakov
suppression is weak at low momentum transfer $Q$, the soft
contribution can not be suppressed effectively. Whether twist-3
contribution is large or small should be checked by other
complementary non-perturbative method in order to guarantee the
validity of twist expansion.

At last, we briefly compare our perturbative analysis with the prediction
from the QCD sum rule. There is a major problem about OPE
(operator product expansion) in the early three-point QCD sum rule
so that QCD sum rule is hold for $Q<M$ where $M$ is the Borel
parameter which is at the order of $1-2\GeV^2$ \cite{Braun}. At this
energy region, the factorization of the hard and soft dynamics is
generally difficult or impossible. In the light-cone sum rules,
the expansion is carried out by twist of hadrons so that comparison
of the perturbative framework and light-cone sum rule becomes possible. A
systematic analysis of pion form factor including the radiative
corrections in the light-cone sum rule approach is provided in
\cite{Braun}. They found that the soft endpiont
contribution is dominant and the hard contribution is very small.
Our analysis also shows LO twist-2 hard contribution is very
small. But the calculation of the hard part of twist-2
contribution is different in these two approaches in principle. In
pQCD approach, the momentum of exchanged gluon is determined by
the longitudinal and transverse momentum of the quarks in the two pions. The
LO calculation is at the tree level. In the light-cone sum rule,
the LO order is soft endpoint contribution and the NLO radiative
corrections contains hard and soft contributions. There is a
certain seeming correspondence between the single-gluon exchange
diagram in pQCD approach and a diagram of radiative corrections in
the light-cone sum rule approach. However, they are not equivalent. In
general, the loop corrections are basically different from tree
level calculation. In loop corrections, the momentum is arbitrary
and the result has ultraviolet divergence and infrared divergence.

In conclusion, in large momentum transfer region, say
$Q^2>20\GeV^2$, non-perturbative contribution is small, EM pion
form factor can be self-consistently calculated by pQCD approach.
While, in small momentum transfer region, $Q^2<10\GeV^2$, Sudakov
suppression becomes weak, non-perturbative contribution can not
be completely suppressed. There is always at least 20\% soft
contribution left in pQCD prediction if we use the soft criterion
$\alpha_s>0.5$. This soft contribution breaks the self-consistence
of pQCD approach. The solution of this problem should be beyond
the perturbative framework. The reliable calculation of power
correction requires non-perturbative methods.

\section*{Acknowledgment}

We would like to thank H. Li for many helpful discussions. Z. Wei
thanks to H. Cheng, G. Sterman, S. Brodsky, and V. Braun for their
discussions and comments and the partial support of National
Science Council of R.O.C. under Grant No. NSC
91-2816-M-001-0004-6. M. Yang  also thanks C.D. l\"{u} for
valuable discussions and the partial support of the Research Fund
for Returned Overseas Chinese Scholars and National Natural
Science Foundation of China.


\begin{thebibliography}{99}

\bibitem{BLER} S.  Brodsky and G.  Lepage, \prl{43}{545}{1979};
   \plb{87}{1979}{359}; \prd{22}{1980}{2157};
  A. Efremov and A. Radyushkin, \plb{94}{1980}{245}.

\bibitem{Kroll} P. Kroll, M. Raulfs, \plb{387}{1996}{848-854}.

\bibitem{Isgur} N. Isgur and C. Smith, \npb{317}{1989}{526-572};
A.V. Radyushkin, \npb{325}{1991}{141}.

\bibitem{LiSterman}  H. Li and G. Sterman, \npb{381}{1992}{129-140}.

\bibitem{pQCD} H. Li and H. Yu, \prl{74}{1995}{4388-4391};
  Y. Keum, H. Li and A.I. Sanda, \plb{504}{2001}{6};
    \prd{63} {2001}{054008};
  C. L\"u, K. Ukai and M. Yang,  \prd{63} {2001}{074009};
  D. Du, C. Huang, Z. Wei and M. Yang, \plb{520}{2001}{50-58};
  C. L\"u and M. Yang, \epjc{23}{2002}{275}.

\bibitem{Sachrajda} S. Descotes and  C.T. Sachrajda, \npb{625}
  {2002}{239-278}.

\bibitem{WeiYang} Z. Wei and M. Yang, \hepph{0202018}.

\bibitem{Braun} V.M. Braun, A. Khodjamirian and M. Maul,
 \prd{61}{2000}{073004}.

\bibitem{Jakob} R. Jakob and P. Kroll, \plb{315}{1993}{463-470}.

\bibitem{Melic} B. Meli$\acute{\rm c}$,
 B. Ni$\breve{\rm z}$i$\acute{\rm c}$
 and K. Passek, \prd{60}{1999}{074004}.

\bibitem{twist-3} B.V. Geshkenbein and M.V. Terentev,
 \npb{117}{1982}{243-246}.

\bibitem{CYH} F. Cao, Y. Dai and C. Huang,
 \epjc{11}{1999}{501-506}.

\bibitem{BottsSterman} J. Botts and G. Sterman, \npb{325}{1989}{62-100}.


\bibitem{Brauntwist} V.M. Braun and I.E. Filyanov,
  \zpc{44}{1989}{157}; \zpc{48}{1990}{239}.



\bibitem{Lithreshold} H. Li,  preprint: \hepph{0102013}.



\end{thebibliography}
\end{document}